\documentclass[twocolumn,superscriptaddress,showpacs,preprintnumbers,amsmath,amssymb]{revtex4-1}
\usepackage{epsfig,rotate,colordvi}
\usepackage{subfigure}
\usepackage{graphicx}

\begin{document}
\title{Gap states in insulating LaMnPO$_{1-x}$F$_{x}$ (x = 0 - 0.3)}

\author{J. W. Simonson}
\email{jsimonson@bnl.gov}
\affiliation{Department of Physics and Astronomy, Stony Brook University, Stony Brook, NY 11794}
\author{K. Post}
\affiliation{Department of Physics, University of California, San Diego, CA 92093}
\author{C. Marques}
\affiliation{Department of Physics and Astronomy, Stony Brook University, Stony Brook, NY 11794}
\author{G. Smith}
\affiliation{Department of Physics and Astronomy, Stony Brook University, Stony Brook, NY 11794}
\author{O. Khatib}
\affiliation{Department of Physics, University of California, San Diego, CA 92093}
\author{D. N. Basov}
\affiliation{Department of Physics, University of California, San Diego, CA 92093}
\author{M. C. Aronson}
\affiliation{Department of Physics and Astronomy, Stony Brook University, Stony Brook, NY 11794}
\affiliation{Condensed Matter Physics and Materials Science Department, Brookhaven National Laboratory, Upton, NY 11793}

\date{\today}

\begin{abstract}
Infrared transmission and electrical resistivity measurements reveal that single crystals of LaMnPO$_{1-x}$F$_{x}$ (x $\leq$ 0.28) are insulating. The optical gap obtained from transmission measurements is nearly unaffected by doping, decreasing only slightly from 1.3 eV in undoped LaMnPO to 1.1 eV for x = 0.04. The activation gaps obtained from electrical resistivity measurements are smaller by at least an order of magnitude, signalling the presence of states within the optical gap.  At low temperatures, the resistivity is described well by variable range hopping conduction between these localized gap states.  Analysis of the hopping conduction suggests that the gap states become slightly more delocalized with fluorine content, although metallic conduction is not observed even for fluorine concentrations as large as x = 0.28.
\end{abstract}

\maketitle
Correlation gap insulators are distinct from simple band insulators in that the former require an electron-electron interaction to open a band gap at the Fermi level.  This interaction typically stems from strong Coulomb repulsion (U) between the charge carriers, which may or may not be accompanied by further restrictions on orbital occupancy that are derived from the often subtle interplay between crystal field splitting and Hund's rules.  These correlations can be weakened by carrier doping or by the application of high pressures~\cite{imada1998}, in either case resulting in the transformation of the correlation gap insulator into a correlated metal at a Mott-like electronic delocalization transition (EDT).  It is believed that the correlation gap itself remains essentially unchanged by this process and that the EDT involves the delocalization of intrinsic states located within the gap~\cite{kotliar2006}.  Initially these states are highly localized by strong Coulomb interactions.  Their density increases as the EDT is approached, however, and a band of correlated states forms as the correlations become weaker.  In this way, the in-gap states found in a correlation gap insulator are the intrinsic precursors of the correlated metal that forms once the EDT occurs, although the gap itself remains almost unchanged.  Qualitatively speaking, this scenario is consistent with the wealth of experimental information available on the cuprates, where charge doping drives the magnetically ordered and insulating parent compounds metallic and ultimately superconducting~\cite{basovrmp}.  

The more recent discovery of high superconducting critical temperatures (T$_{c}$) in the iron pnictide family, most notably T$_{c}$ = 55 K in F-doped SmFeAsO with the ZrCuSiAs structure~\cite{ren2008}, often draws comparison to the cuprate superconductors~\cite{qazilbash2009,basov2011}.  Nearly all compounds of the iron pnictide family so far reported are metals, however, falling on the itinerant side of a possible EDT.  Two insulating exceptions have been reported, but neither provides the perfect analogue to La$_{2}$CuO$_{4}$: (K,Tl)Fe$_x$Se$_2$~\cite{fang2011} is subject to phase segregation and chemical inhomogeneity~\cite{song2011}, while a doping study of compounds within the R$_2$O$_2$Fe$_2$O(Se, S)$_2$ (R = La, Ce, Pr, Nd, Sm) system~\cite{zhu2010,ni2011} has not yet been undertaken.  Nonetheless, theoretical studies have suggested that the Mott physics embodied in the cuprates remains relevant in the iron pnictides~\cite{si2008,kotliar2008}.  Can the familiar theme of doping a Mott insulator across an EDT also be realized in the iron pnictide/chalcogenide family?  If so, the presence of heightened correlations near the transition might lead to further enhancement of the maximum T$_{c}$ in this class of superconductors~\cite{qazilbash2009,basov2011}.

Unlike the iron pnictides, the layered manganese pnictide compounds are often insulators, particularly those compounds for which the formal manganese valence is two.  This tendency is realized in the so-called 122 compounds like BaMn$_2$Sb$_2$~\cite{wang2009}, the 1111 compounds like LaMnPnO (Pn = P, As, Sb)~\cite{kayanuma2009}, and the 111 compounds such as LiMnAs ~\cite{jungwirth2011}.  Studies of the effects of doping in compounds of this type, however, are so far very limited.  In one instance the properties of polycrystalline LaMnPO were investigated, and that compound was found to remain an insulator when doped with up to 10\% copper to the manganese site or 7\% calcium to the lanthanum site~\cite{yanagi2009}.  Unfortunately, the inherent compositional variability within the polycrystalline matrix also produced room temperature resistivities that varied by as much as two orders of magnitude between nominally identical undoped samples.  Further studies carried out on polycrystalline RMnPO (R = Nd, Sm, Gd) with 3\% calcium on the R site or similar quantities of Cu on the Mn site found that the electrical resistivity of the undoped compounds was several orders of magnitude lower than that of LaMnPO.  Otherwise, the results were qualitatively similar in that no EDT was observed ~\cite{yanagi2010}.  Remarkably, an EDT has been very recently reported in polycrystalline SmMnAsO$_{1-x}$ samples prepared under high pressure that were effectively electron doped via oxygen deficiency~\cite{shiomi2011}.  Here, the EDT is accompanied by a substantial decrease in the N\'{e}el temperature T$_{N}$.  The presence of a 4\% volume fraction of SmAs was also observed, however, leading to an increase in the specific heat below 3 K, and there are unmistakeable signatures in the magnetic susceptibility of ferromagnetic MnAs below 320 K, highlighting the inherent difficulties of working with polycrystalline samples, while suggesting that further investigation may be necessary.

In light of both the overall paucity of published data on doped layered manganese pnictides and the fundamental irreproducibility of measurements of physical properties obtained from polycrystalline samples, we undertook a systematic study of the transport properties of fluorine-doped LaMnPO single crystals with the objective of determining whether an electronic delocalization transition occurs with the introduction of charged impurities.  We selected LaMnPO for this study because it forms in the same ZrCuSiAs structure as the 26 K superconductor LaFeAsO~\cite{kamihara2008}.  Further, we note that LaMnPO is not likely to be a simple band insulator, given its 21 valence electrons.  Formal valence counting suggests a half-filled d$^5$ configuration for the manganese atoms, and DFT calculations performed in the absence of magnetic order have predicted LaMnPO to be a metal~\cite{xu2008}. This result suggests that its insulating nature is the result of correlations associated with the magnetism of the manganese atoms.  Although we doped this compound across a larger electron range than had been accessed previously, we found that both the transport properties and the optical gap changed relatively little with charge doping.  Nonetheless, analysis of the low temperature conductivity reveals that the system is indeed driven somewhat closer to delocalization with the addition of fluorine as an electron dopant, although metallic conduction is never achieved.

\begin{table*}
\caption{\label{tab:table1}RefFIT parameters of position ($\omega_o$), strength ($\omega_p$) and broadness ($\gamma$) corresponding to Lorentzian oscillators for the optical gap and intragap absorption.  The quantity d is the sample thickness as determined by the model.}
\begin{ruledtabular}
\begin{tabular}{cccccc}
 LaMnPO$_{1-x}$F$_{x}$&$\omega_o$ (cm$^{-1}$)&$\omega_p$ (cm$^{-1}$)&$\gamma$ (cm$^{-1}$)&d ($\mu$m)\\ \hline
 x = 0   &1703.2&1169.7&9450.5&6.8835\\
         &9429.3&1231.9&688.1& \\
 x = 0.02&1546.5&1180.8&6029.8&13.485\\
         &8867.4&1231.9&481.71& \\
 x = 0.03&1037.0&2202.4&4913.4&8.3274\\
         &8867..4&1574.6&613.89& \\
 x = 0.04&1331.0&1620.9&4925.3&6.4104\\
         &9133.4&1767.2&568.42& \\
 \end{tabular}
\end{ruledtabular}
\end{table*}
We grew plate-like single crystals of LaMnPO$_{1-x}$F$_{x}$ (0 $\leq$ x $\leq$ 0.28) that were as large as 2 x 3 x 0.01 mm from a NaCl-KCl eutectic flux, following an established method~\cite{nientiedt1997}.  The crystal structure was confirmed using a Bruker Apex II single crystal x-ray diffractometer, and x-ray diffraction performed on powdered crystals found no evidence of secondary phases.  Room temperature infrared (IR) transmission spectra were measured using a Bruker Vertex v/70 FT-IR spectrometer coupled to an IR microscope, which allowed us to obtain reliable data even on small ($<$ 1 mm$^2$) crystals.  Both AC and DC electrical resistance measurements were carried out in a Quantum Design Physical Properties Measurement System on single crystals with current flowing perpendicular to the crystallographic c direction.  Given the wide range of resistance values among the doped crystals, excitation currents were chosen from values between 95 nA and 100 $\mu$A in order to minimize heating of the samples during measurement.  Fluorine concentrations were measured via the potentiometric method using a LaF crystal as an ionic filter ~\cite{frant1966}.  LaMnPO single crystals of known mass were dissolved in aqua regia, which was then buffered and decomplexed with the addition of controlled amounts of total ionic strength adjustment buffer IV (TISAB IV) to the solution.  The potentiometric measurements were standardized to absolute fluorine concentrations that were obtained from pyrohydrolysis measurements on polycrystalline samples (ATI Wah Chang, Albany, OR).

\begin{figure}
\includegraphics[width=6.5cm]{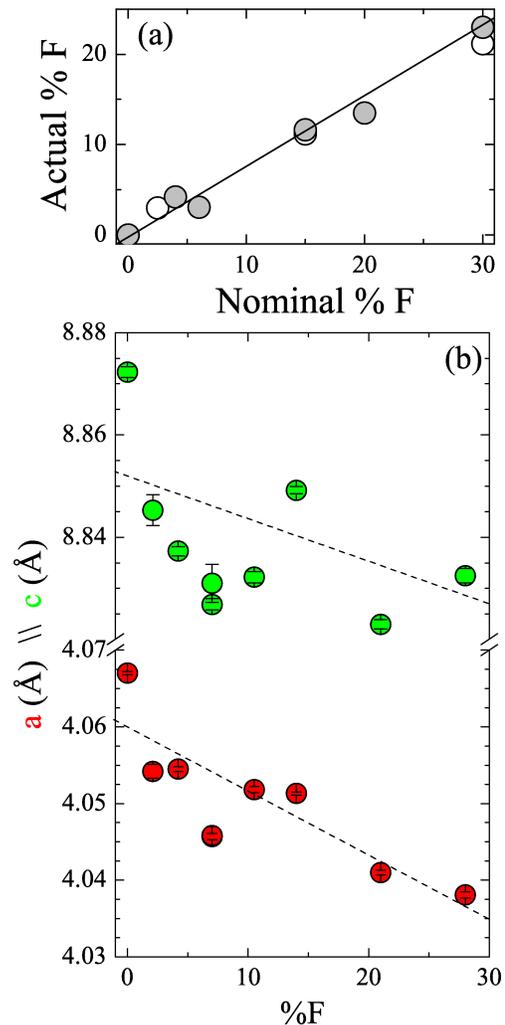}
\caption{ (Color online) (a) The actual and nominal fluorine concentrations in LaMnPO$_{1-x}$F$_{x}$ are linearly related, both for single crystals (shaded circles) and for polycrystalline standards (open circles). The solid line indicates a linear fit with slope 0.71$\pm$0.04. (b) The  a (red points) and c (green points) lattice parameters both decrease linearly with actual fluorine concentration. Dashed lines are guides for the eye. }
\end{figure}
Fig. 1 provides confirmation that we were successful in incorporating fluorine into the LaMnPO structure and in quantifying its concentration.  Fig. 1a plots the actual fluorine composition measured using an ion selective electrode as a function of the nominal fluorine concentration.  The slope of the line indicates that about 71 $\pm $4$\%$ of the fluorine introduced  in the growth was ultimately incorporated in the single crystals. There is no indication of a solubility limit for fluorine in LaMnPO, although we were not able to grow crystals via this approach with x $\geq$ 0.3.  The compositions discussed in this article always refer to the actual values of fluorine concentration.  Measurements of the lattice constants determined by single crystal x-ray diffraction are plotted in Fig. 1b.   With the addition of 28 $\pm$ 2\% fluorine, the intralayer $a$ lattice constant decreased by $\simeq$0.7\%, while the interlayer $c$ lattice constant was reduced by about 0.4\%.  The sensitivities of the lattice constants to fluorine doping are similar in LaMnPO and in CeFeAsO~\cite{zhao2008}, although the results in that work refer to nominal concentrations.

\begin{figure}
\includegraphics[width=8cm]{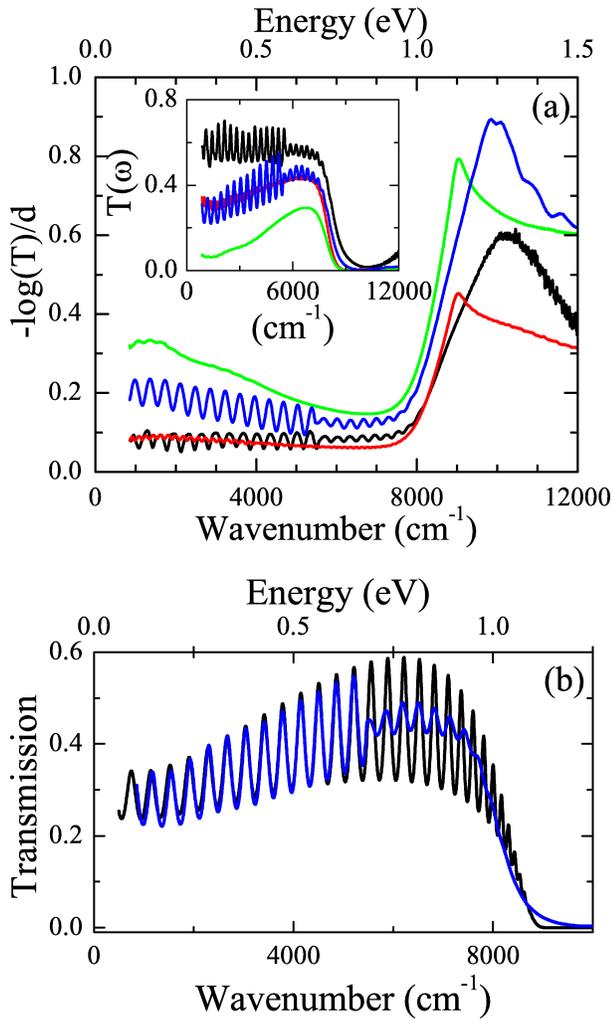}
\caption{(color online) (a) Absorption spectra for LaMnPO$_{1-x}$F$_{x}$, with x = 0 (black), 0.02 (red), 0.03 (green), and 0.04 (blue).  The large peak above 7700 cm$^{-1}$ indicates the onset of the optical gap.  The increase at lower frequencies shows the enhanced intragap absorption with doping.  Inset: absolute transmission spectra.  (b) Comparison of the RefFIT model (black) and the experimental data (blue).  Note the change in amplitude of the experimental interference patterns around 5500 cm$^{-1}$.  This suppression begins at the point where the MIR and NIR data sets are merged and results from a change in spectrometer resolution between the two regions.}
\end{figure}
Fig. 2 shows that the compounds LaMnPO$_{1-x}$F$_{x}$ are insulators.   We used IR spectroscopy to measure the size of the optical gap of LaMnPO and to explore possible changes to the electronic structure that resulted from fluorine doping.  Absolute transmission (T($\omega$)) spectra for pristine LaMnPO and for doped LaMnPO$_{1-x}$F$_{x}$ are shown in the inset to Fig. 2a with x = 0.02, 0.03, and 0.04.  The dominant feature of these spectra is the rapid drop in transmission above 8000 cm$^{-1}$ originating from transitions across the insulating gap that are direct with respect to momentum.  In order to quantify the magnitude of the energy gap, it is instructive to display these data in the form of absorption, A = -log(T)/d, where d is the sample thickness.  To extract the thickness of the crystals, experimental T($\omega$) spectra were modeled using RefFIT software~\cite{kuzmenko2004} in order to match closely both the absolute values of transmission and the interference patterns at low frequencies.  We used two Lorentzian oscillators within the Drude-Lorentz model such that:
\begin{equation}\epsilon(\omega) = \epsilon_{\infty} + \sum_{i}\frac{\omega_{pi}^2}{\omega_{oi}^2-\omega^2-i\gamma_i\omega} \label{eq:DL}\end{equation}
where $\epsilon(\omega)$ is the dielectric function and $\omega_{pi}$, $\omega_{oi}$, and $\gamma_i$, are respectively the plasma frequency, eigenfrequency, and linewidth of the $i$th Lorentzian oscillator.  The first of the two components of our model is a strong oscillator corresponding to the optical gap, while for the second we included a broad, weaker oscillator at lower wavenumber to account for intragap absorption.  By varying the parameters of these two oscillators, as well as the sample thickness, we were able to reproduce the experimental results closely and to obtain a reasonable value of the sample thickness.  The values for the parameters are listed in table I, and an example of this fitting is shown for x = 0.04 in Fig. 2b.

The absorption spectra are displayed in Fig. 2a and are dominated by an absorption peak at 8000 cm$^{-1}$.  There is a noticeable suppression of the interference patterns around 4500 cm$^{-1}$ that is due to a change in spectrometer resolution between the mid-infrared (MIR) and near-infrared (NIR) experiments and is not an intrinsic property of the sample.  Therefore, we attempted to reproduce only the interference patterns in the MIR range in our model.  We notice that it is the period of the interference fringes, and not their amplitude, that allows us to evaluate the sample thickness from IR data.  Although the interference patterns used for modeling cannot always be readily seen in the spectra shown in Fig. 2a (e.g. x = 0.03), they can be clearly resolved through close examination of the data below 8000 cm$^{-1}$.  We also inferred that the dielectric constant ($\epsilon_{\infty}$) changes very little in these samples, so it was kept constant as the other parameters were varied about the values determined for the undoped sample.  This process should produce values of thickness with the correct ratios so that the absorption plot is qualitatively correct.

We determined the optical gap of these crystals from the sharp upturn in the absorption spectra shown in Fig. 2a.  This upturn, which is an attribute of the direct energy gap, occurs around ~7700 cm$^{-1}$ in all samples.  The optical gaps fall within a range of 1.1 to 1.3 eV, with the undoped compound exhibiting the largest value.  This result is in good agreement with a previously reported measurement of the optical gap of undoped LaMnPO obtained from the diffuse reflectance spectrum of a polycrystalline pellet~\cite{yanagi2009}.  Notably, the optical gap of our single crystals is nearly invariant among all samples containing up to 4\% fluorine.  This observation reveals the resilience of the optical gap to dopant concentration.  There is also a significant increase in the background absorption of more highly doped samples (e.g. x = 0.03 and 0.04).  This is an expected result of enhanced intragap absorption with doping, which is consistent with the decrease in room temperature resistivity that will be discussed below.  Attempts were made to measure IR transmission through crystals doped with 10\% and 20\% fluorine, but these crystals were found to be optically opaque, likely a result of their higher fluorine concentrations, precluding further measurement.

\begin{figure}
\includegraphics[width=8.5 cm]{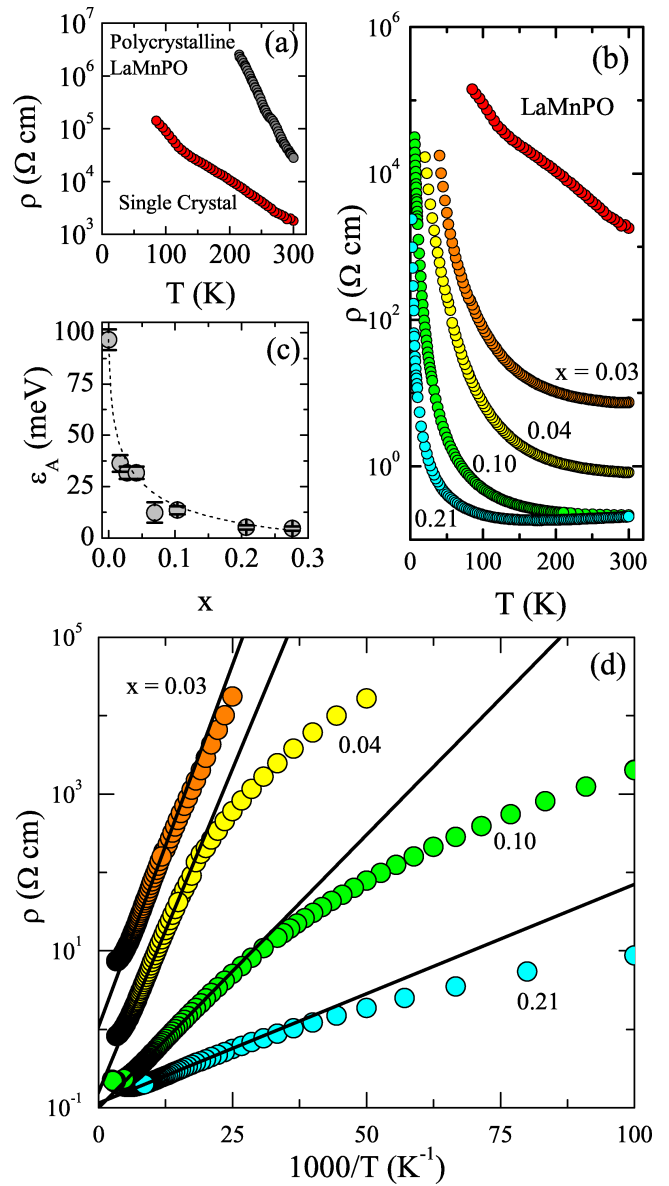}
\caption{(a) Comparison of the temperature dependencies of the electrical resistivities $\rho$ of polycrystalline and single crystal LaMnPO, as indicated. (b) Temperature dependencies of the electrical resistivities of LaMnPO$_{1-x}$F$_{x}$ for different values of x, as indicated. (c) Activation gap $\epsilon_{act}$ as a function of x.  The dashed line is a guide for the eye.  (d) Arrhenius plots of the temperature dependencies of the electrical resistivity for different x, as indicated. Solid lines are best high temperature fits.}
\end{figure}
The activation gap found in resistivity measurements (Fig. 3) is not only much smaller than the optical gap but is also significantly affected by the presence of fluorine.  In Fig. 3a, we compare the resistivity measured in a polycrystalline sample with that of a single crystal to reinforce the point that single crystals are often necessary for the accurate determination of intrinsic transport properties.  The polycrystalline data in the figure are in good agreement with those from the previous report~\cite{yanagi2009}, but the magnitude of the resistivity is much lower for the single crystal, underscoring the pejorative effect that boundary scattering in polycrystalline samples can have on electrical transport.  Fig. 3b displays the electrical resistivity of LaMnPO$_{1-x}$F$_{x}$ single crystals measured at temperatures below 300 K.  Undoped LaMnPO exhibits a high resistivity at room temperature, and the resistivity increases as temperature is lowered, confirming that the compound is an insulator.  This result is in accordance with the conclusions of the optical transmission measurements.  We observed that doping LaMnPO with fluorine rapidly and systematically decreases the resistivity at 300 K by as much as four orders of magnitude when x $\ge$ 0.1, again in agreement with the results of the transmission measurements.  Nonetheless, even for fluorine concentrations as high as x = 0.28, the resistivity increases sharply near the lowest temperatures at which measurement is possible, indicating that the system has not become metallic.  Fig. 3c is a plot of the activation gaps $\epsilon_{act}$ at different fluorine concentrations,  taken from  Arrhenius plots in which the resistivity takes the form $\rho \propto exp(\epsilon_{act}/2k_BT)$. Unlike the optical gap, which is almost independent of fluorine concentration, the activation gap rapidly falls off with dopant content from an initial value of nearly 100 meV in undoped LaMnPO to 5 meV in the most highly doped samples.  Fig. 3d displays Arrhenius plots of the resistivity data along with linear fits to activated conduction.  The resistivity data are well described by an activated expression to temperatures as low as 100 K, while the plots acquire curvature towards lower resistivity at lower temperatures.  Overall, the resistance measurements show that the activation gap in LaMnPO is significantly smaller than the optical gap and that its magnitude decreases with fluorine doping.  These observations indicate that conductivity stems from the excitation of charge carriers from gap states that remain localized even for large fluorine levels.  Nonetheless, the average energy of these excitations is systematically reduced with doping.

Fig. 4 shows that the conductivity at low temperatures is dominated by hopping between localized gap states.  The nonlinearity present in the Arrhenius plots of Fig. 3d is due to variable range hopping (VRH) conduction between in-gap states that emerges as the dominant contribution to the resistivity at the lowest temperatures.  The probability of an electron hopping from one localized state to another is given by the following relation ~\cite{mott1979}:
\begin{equation}P_{hop} \propto exp(-2\alpha r + \frac{\epsilon_{hop}}{k_B T})\label{eq:prob}\end{equation}
where $\alpha$ is the reciprocal of the electron localization length, r is the real space distance between states, and $\epsilon_{hop}$ is the energy difference between the states,  therefore falling on a scale that is less than that of the activation energy, i.e. $\epsilon_{hop} \le \epsilon_{act}$.  In three dimensions, the maximization of equation~\ref{eq:prob} with respect to the hopping distance leads to an electrical resistivity $\rho \propto exp(T_0/T)^{-1/4}$, where $T_0$ parameterizes the disorder in the system.  Fig. 4a shows the resistivity plotted versus $T^{-1/4}$ on a logarithmic scale.  At low temperatures the curves become linear, and the slope yields $T_0$.  We observe that the resistivity in this temperature range is well characterized by 3D VRH conduction and further that the slope $T_0$ decreases systematically with fluorine concentration.  We deduce that these local states are present for all levels of doping and furthermore that there exists a quantitative enhancement of carrier transport with fluorine content that persists to the lowest temperatures, suggesting a change in the nature of the low energy gap excitations.
\begin{figure}
\includegraphics[width=8.5cm]{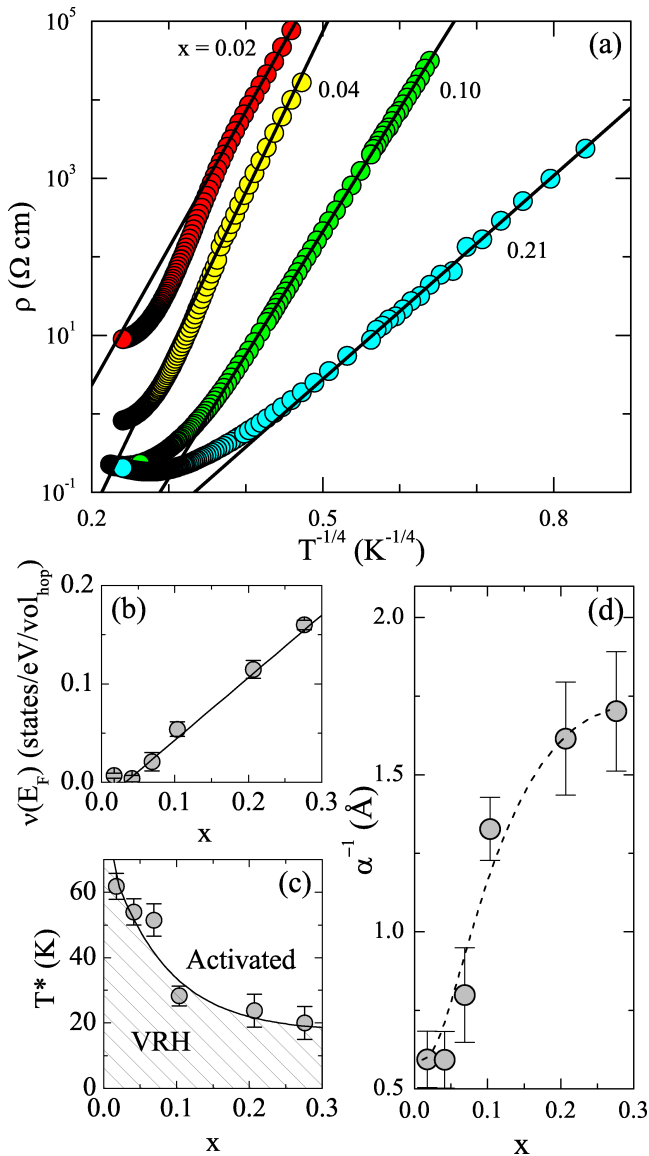}
\caption{ (Color online) (a) Electrical resistivities of LaMnPO$_{1-x}$F$_{x}$  for different x (as indicated) plotted versus $T^{-1/4}$ to display the 3D variable range hopping (VRH) conduction. Solid lines indicate the best regression fits.  (b) The density of states per eV per hopping volume $\nu(E_F)$ as a function of fluorine concentration x. Solid line is guide for the eye. (c) The fluorine concentration x dependence of the crossover temperature T* that separates low temperature VRH from high temperature activated behaviors. Solid line is a guide for the eye.  (d) The localization length $\alpha^{-1}$ (see text) increases with fluorine concentration x. Dashed line is a guide for the eye.}
\end{figure}

Analysis of the variable range hopping data in Fig. 4 allows us to probe more closely the nature of the gap and the states within it.  We define the density of states per hopping volume as $\nu$(E$_F$) = N(E$_F$)/$\alpha^3$, where N(E$_F$) is given in states/eV/$\AA^3$ and $\alpha$ is the electron localization parameter from equation~\ref{eq:prob}. Then the expression for the density of states at E$_F$ in a VRH system becomes $\nu$(E$_F$) = 1.5/k$_B$T$_0$, where $T_0$ is the disorder parameter determined in Fig. 4a and we take Mott's numerical parameter of 1.5~\cite{mott1979}.  This quantity, corresponding to a density of states per localization or hopping volume, is plotted in Fig. 4b and increases consistently with the percentage of fluorine doped on the oxygen site.  The relative enhancement of $\nu$(E$_F$) surpasses an order of magnitude for the most highly doped crystals.  It can be inferred from our definition of $\nu$(E$_F$) that this increase must stem from one of two origins: either from an augmentation of the in-gap density of localized states N(E$_F$) as fluorine is added to the structure or from a decrease in the localization parameter that is indicative of a systematic delocalization of the gap states.

We resolve the question of whether localization decreases with fluorine doping or alternatively if this enhancement stems solely from an increase in N(E$_F$) by defining T* as the crossover temperature, above which VRH conduction gives way to activated behavior due to the dominance of nearest neighbor hopping.  Specifically, T* was determined as the temperature at which the regression fits of the VRH and activated modes of conduction intersect one another.  The trend in T* with x is plotted in Fig. 4c, which shows that this crossover temperature decreases systematically with fluorine content, consistent with the concurrent reduction in the size of the activation gap.  The localization parameter $\alpha$ can then be isolated by again maximizing equation~\ref{eq:prob} with respect to hopping distance and by setting the proper boundary conditions for nearest neighbor hopping.  These conditions imply that $\epsilon_{hop}$ is given by the activation energy determined from the Arrenhius plots, T = T*, and r = r$_{nn}$ is the nearest neighbor distance as determined by x-ray diffraction.  The electron localization length is then given by the quantity 1/$\alpha$ = r$_{nn}$(2k$_B$T*)/$\epsilon_{act}$, which we plot in Fig. 4d for fluorine-doped LaMnPO.  This length increases by more than a factor of three  in LaMnP(O$_{1-x}$F$_{x}$) as the fluorine concentration is increased from zero to x = 0.3, indicating growth of the spatial extent of the localized states within the gap.  The ramifications of this result are that not only is the activation barrier reduced with fluorine doping, but also that the states pinning E$_F$ within the gap are systematically pushed towards delocalization, even if the compound as a whole fails to become metallic.

Overall these experiments reveal two key points about LaMnPO: that the size of the optical gap does not change with fluorine concentration and that in-gap states persist and remain localized even at the highest fluorine concentrations.  The relative insensitivity of the optical gap to doping is expected for both correlation gap and conventional insulators.  The in-gap states in LaMnPO, however, are remarkably insensitive to doping.  For instance, introducing charge into only a few percent of  all unit cells  transforms cuprates such as Nd$_{2-x}$Ce$_x$CuO$_4$ and La$_2$CuO$_4$ from moment bearing and antiferromagnetically ordered insulators into strongly correlated metals\cite{budnick1988}.  Although there is no insulating parent compound in the iron pnictides, a truncated version of the same scenario is realized here, with 10-15 $\%$ charge doping required to induce superconductivity in systems like LaFeAs(O$_{1-x}$F$_{x}$)~\cite{kamihara2008}.  While the reduction of the resistivity and the diminishment of the activation gap suggest that LaMnPO is slowly evolving towards metallization, the in-gap states remain restricted to length scales that are smaller than the unit cell, even as the number of doped charges approaches one per formula unit. For a conventional insulator, this would induce an extremely large degree of disorder, likely much more than our structural refinements of LaMnPO would allow.  We suggest instead that the unusually large degree of electronic localization that our experiments reveal in LaMnP(O$_{1-x}$F$_{x}$) may result from the large Hund's rule interactions characteristic of Mn$^{2+}$, which can be expected to be largely impervious to changes in electron concentration if the Mn valence is left unaffected.  Simply put, the local in-gap states seem to be just as correlated as their band-like relatives in undoped and insulating LaMnPO.

We have presented optical transmission and electrical resistivity measurements of LaMnPO doped with fluorine.  The optical gap was measured to decrease no more than $\simeq$ 10\% with fluorine concentrations as large as 4\%, consistent with the theoretical picture of the EDT in correlation gap insulators.  The activation gap determined from electrical resistivity measurements is smaller than the optical gap by an order of magnitude, and decreases strongly with fluorine doping.  We observed that fluorine doping falls short of  metallizing LaMnPO even at high concentrations, in stark contrast to the case of La$_2$CuO$_4$ and other high temperature superconductors.  These results indicate that carrier transport in LaMnPO is dominated by the presence of localized states that pin the Fermi energy within the gap, and no metallic conduction is observed for fluorine concentrations as large as x = 0.3.  The low temperature transport in particular is carried out via variable range hopping, and analysis of this hopping conduction suggests that fluorine doping promotes a tendency towards greater delocalization of the gap states.  LaMnPO never reaches the EDT, however, because the gap states remain localized, never broadening enough to form the delocalized impurity band that is observed in the cuprate and iron pnictide superconductors.  While LaMnPO may be too strongly correlated to be doped across an EDT, other layered manganese pnictides may be more amenable to inducing an EDT via charge doping, leaving open the door to the formation of strongly correlated metals and even superconductors in this family of compounds.

This work was carried out under the auspices of a Department of Defense National Security Science and Engineering Faculty Fellowship via Air Force Office of Scientific Research grant FA 9550-10-1-0191.  O. Khatib was supported by the Department of Energy, Basic Energy Sciences.

\end{document}